\documentclass[journal=jacsat, manuscript=article, layout=twocolumn]{achemso}
\usepackage[version=3]{mhchem}
\usepackage[utf8]{inputenc}
\usepackage[T1]{fontenc}
\usepackage{amssymb}
\usepackage{hyperref}
\usepackage{tikz}
\usepackage{siunitx}
\usepackage{placeins}
\DeclareSIUnit\calorie{cal}
\usepackage{lineno}

\newcommand{\reviewerchanges}[1]{#1}

\emergencystretch=\maxdimen
\author{Lukas H\"ormann}
\email{hoermann@tugraz.at}
\affiliation[TUG]
{Institute of Solid State Physics, Graz University of Technology, Graz}

\author{Johannes J. Cartus}
\affiliation[TUG]
{Institute of Solid State Physics, Graz University of Technology, Graz}

\author{Oliver T. Hofmann}
\affiliation[TUG]
{Institute of Solid State Physics, Graz University of Technology, Graz}

\title{The impact of static distortion waves on superlubricity}

\keywords{Superlubrictiy, Static distortion waves, Incommensurate structures, Organic/metal interfaces, Machine learned interatomic potentials}

\let\oldmaketitle\maketitle
\let\maketitle\relax

\begin{document}

\begin{tocentry}
\includegraphics[width=\linewidth]{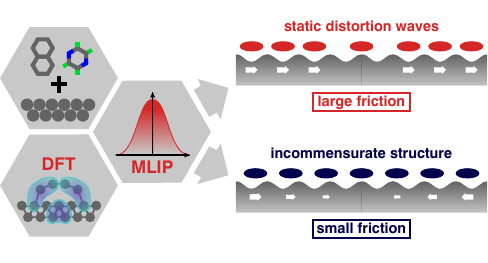}
\end{tocentry}

\twocolumn[
\begin{@twocolumnfalse}
\oldmaketitle
\begin{abstract}
Friction is a major source of energy loss in mechanical devices. This energy loss may be minimized by creating interfaces with extremely reduced friction, i.e. superlubricity. Conventional wisdom holds that incommensurate interface structures facilitate superlubricity. Accurately describing friction necessitates precise modeling of the interface structure. This, in turn, requires the use of accurate first-principles electronic structure methods, especially when studying organic/metal interfaces, which are highly relevant due to their tunability and propensity to form incommensurate structures. However, the system size required to calculate incommensurate structures renders such calculations intractable. As a result, studies of incommensurate interfaces have been limited to very simple model systems or strongly simplified methodology. We overcome this limitation by developing a machine-learned interatomic potential that is able to determine energies and forces for structures containing thousands to tens of thousands of atoms with an accuracy comparable to conventional first principles methods but at a fraction of the cost. Using this approach, we quantify the breakdown of superlubricity in incommensurate structures due to the formation of static distortion waves. Moreover, we extract design principles to engineer incommensurate interface systems where the formation of static distortion waves is suppressed, which facilitates low friction coefficients.
\end{abstract}
\end{@twocolumnfalse}
]

\section{Introduction}

Friction causes a significant amount of energy consumption in any moving mechanical device. One way to minimize this energy loss is creating interfaces with extremely reduced friction, i.e. superlubricity. Superlubricity is a state of ultra-low friction, defined by a dynamic friction coefficient below \num{0.01}.\cite{baykara2018emerging, zhai2019nanomaterials} \reviewerchanges{Such low friction coefficients can be realized in form of liquid superlubircity,\cite{han2022unlocking, zhai2019nanomaterials} using control or actuation of the normal force,\cite{PhysRevLett.92.134301, PhysRevLett.97.136106, socoliuc2006atomic, gnecco2008dynamic, ma2023active, shi2021micro} thermally activated drift,\cite{PhysRevE.71.065101, PhysRevB.78.155440, PhysRevB.79.045414, PhysRevB.81.245415, PhysRevLett.92.126101} or structural superlubricity.\cite{liu2012observation, zhang2013superlubricity, song2018robust, cihan2016structural, benassi2015breakdown, ma2015critical, sharp2016elasticity, berman2015macroscale, berman2018operando, liu2017robust}}

Structural superlubricity is one of the most promising strategies to achieve ultra-low friction.\cite{hirano1990atomistic, shinjo1993dynamics, baykara2018emerging} This type of superlubricity relies on structural incommensurability, where the two surfaces (of a substrate and an adsorbate) that meet at the sliding interface exhibit sightly different lattice spacings. Using the Frenkel-Kontorova model,\cite{kontorova1938k} it can be shown that for a surface with a lattice spacing of $a$ and a rigid adsorbate with a lattice spacing of $b$ lateral forces become infinitesimal beyond a critical value of $a/b$.\cite{peyrard1983critical} However, incommensurate structures are often subject to deformations that may be driven by the forces occurring during interfacial sliding,\cite{filippov2008torque, PhysRevE.81.046105} or are a result of phase transitions,\cite{hormann2022bistable, panich2012phase, stephens1984high, kilian2008commensurate} or static distortion waves.\cite{novaco1977orientational, novaco1979theory, mctague1979substrate, novaco1980theory, meissner2016flexible, forker2017classification} The precise quantification of the impact these deformations have on the frictional properties is of great interest for the design of superlubricating interfaces.

Earlier studies have demonstrated a breakdown of the superlubric state of an incommensurate configuration of a graphene flake on graphite as a result of a rearrangement into a commensurate structure.\cite{PhysRevE.81.046105} Moreover, the impact of the sliding speed, sliding direction, temperature, and normal force has been studied for the same system.\cite{PhysRevLett.100.046102} However, both studies concentrated on a system consisting purely of carbon atoms and employed classical force field potentials. In fact, many theoretical works on nanoscale friction have focused on idealized systems and small unit cells,\cite{wang2012theoretical} which are too small to capture mesoscale phenomena such as static distortion waves. Conversely, investigations of larger or more complex systems have largely used classical force field methods,\cite{samadashvili2013atomistic, falk2010molecular} which do not account for quantum mechanical effects at the interface, such as interfacical charge transfer or hybridization.\cite{hofmann2021first, maurer2019advances} These quantum mechanical effects impact the balance of interactions at the interface that, for instance, governs whether a commensurate or incommensurate structure forms.\cite{hooks2001epitaxy, mannsfeld2005understanding} Therefore, structure determination necessitates first principles electronic structure methods, such as density functional theory (DFT). Consequently, the determination for energies and forces required to calculate friction should be done using the same level of theory. Due to their computational cost, first principles methods can describe continuous interface structures only by using periodic boundary conditions with small unit cells (containing only one or at most a few molecules) which, in practice, restricts these methods to commensurate structures. Conversely, static distortion waves exhibit periodicities that stretch across large unit cells containing hundreds of molecules.\cite{forker2017classification, meissner2016flexible} This large system size renders first principles computation for incommensurate interfaces intractable. We overcome these limitations by developing a machine-learned interatomic potential (MLIP) for pseudo-incommensurate organic/metal interface systems (containing hundreds of molecules per unit cell). This MLIP is trained on state-of-the-art DFT computations allowing us to account for quantum mechanical effects in systems that contain thousands to tens of thousands of atoms. By considering interfaces of this scale we push the boundaries of first principles studies on nanoscale friction to the previously inaccessible mesoscale.

Using this MLIP, we investigate the static friction coefficient of interfaces between metal substrates and molecular adlayers (organic/metal interfaces). At the nanoscale, the dynamic friction coefficient (which is commonly used to define superlubricity\cite{baykara2018emerging, zhai2019nanomaterials}) and the static friction coefficient are related in the following way: During sliding, the surface atoms must overcome energy barriers, distorting them from their equilibrium positions. The maximal lateral force that is required to overcome the barrier yields the static friction coefficient. Once a barrier is overcome, the distortions relax, whereby mechanical energy is dissipated (converted into other forms of energy).\cite{park2014fundamental, berman2018approaches} The amount of dissipated energy for a given sliding distance results in the dynamic friction coefficient. The static friction coefficient is usually larger than the dynamic friction coefficient,\cite{costagliola2016static, huang2020coefficients, barrett1990fastener} which allows us to draw conclusions on superlubricity based on the static friction coefficient. We thoroughly quantify the dependence of the static friction coefficient on the type of commensurabilty as well as static distortion waves by focusing on two exemplary organic/metal interfaces: Naphthalene on Cu(111) and tetrachloropyrazine (TCP) on Pt(111). Naphthalene is a simple polycyclic aromatic hydrocarbon. It forms a variety of physisorbed adlayer structures on the Cu(111) surface. In experiment\cite{yamada2010novel, forker2014complex} and theory\cite{hormann2019sample} this system exhibits commensurate and incommensurate structures at various thermodynamic conditions. Moreover, there is experimental indication that naphthalene on Cu(111) forms static distortion waves.\cite{forker2014complex} TCP is a heterocyclic aromatic organic molecule. On the Pt(111) surface TCP forms chemisorbed and physisorbed adsorption states.\cite{liu2013molecular} In an earlier work\cite{hormann2022bistable} we found strong evidence that chemisorbed TCP forms commensurate and physisorbed TCP forms incommensurate structures. The fact that both systems exhibit both commensurate and incommensurate structures, including static distortion waves in case of naphthalene on Cu(111) allows us to study the dependence of the frictional properties on static distortion waves.



\section{Methods}

\paragraph{Machine-learned interatomic potential for incommensurate structures.} The study of the static friction coefficient of incommensurate interface structures requires a number of energy (and force) evaluations for systems containing hundreds of molecules per unit cell. We determine the formation energy $E_\mathrm{form}$ of a structure from the difference between the energy of the adsorbed system $E_\mathrm{sys}$ and the energy of the pristine substrate $E_\mathrm{sub}$ as well as the energy of the relaxed molecule in vacuum $E_\mathrm{mol}$ multiplied by the number of molecules in the unit cell $N$.

\begin{equation}
	E_\mathrm{form} = E_\mathrm{sys} - E_\mathrm{sub} - N \cdot E_\mathrm{mol}
\end{equation}

The high computational cost associated with calculating these energies prohibits the sole use of first-principles calculations. Therefore, we employ an MLIP based on Gaussian process regression (GPR) and the smooth-overlap-of-atomic-positions descriptor (SOAP)\cite{bartok2013representing} to efficiently determine the energies and forces that are necessary to calculate the static friction coefficient. \reviewerchanges{Combinations of GPR and SOAP have been highly successful for applications including small organic molecules, molecular crystals, metals and semiconductors.\cite{PhysRevLett.104.136403, PhysRevB.95.094203, de2016comparing, PhysRevB.88.054104, PhysRevB.90.104108, bartok2017machine, musil2018machine}} To train these MLIPs we use the energies and forces determined with first principles computations. Separate \reviewerchanges{MLIPs} are used for the individual systems: (A) Naphthalene on Cu(111), (B) physisorbed TCP on Pt(111) and (C) chemisorbed TCP on Pt(111). All MLIPs are trained to achieve a leave-one-out-cross-validation-room-mean-square-error (LOOCV-RMSE) that is smaller than chemical accuracy, i.e. \SI{40}{\milli\electronvolt} per molecule or \SI{1}{\kilo\calorie\per\mol} on the training set. The LOOCV-RMSEs for all MLIPs are shown in the Supporting Information.

To improve the efficiency of the approach we employ a two-step approach: In the first step, we determine an approximate model for the formation energy $E_\mathrm{form}^\mathrm{approx}$. We split this approximate formation energy into a contribution from the molecule-substrate interaction and one from the molecule-molecule interaction:

\begin{equation}
	E_\mathrm{form}^\mathrm{approx} = \sum_i E_\mathrm{mol-sub}^i + E_\mathrm{mol-mol}
\end{equation}

The molecule-substrate interaction is calculated from the sum of molecule-substrate interactions of an isolated molecule on the surface $E_\mathrm{mol-sub}^i$. To determine $E_\mathrm{mol-sub}^i$ we train an MLIP to predict the potential energy surface (PES) of a single isolated molecule on the substrate (see figure \ref{fig:figure1}). For the molecule-molecule interactions $E_\mathrm{mol-mol}$ we use two different approaches. In the case of physisorbed TCP on Pt(111), we train an MLIP on energies of free-standing layers of molecule in vacuum (i.e. adlayer structures whose substrate was removed). This gives an approximate value for the molecule-molecule interaction (often called monolayer formation energy). This allows us to produce training data for large unit cells which could not be calculated if the substrate was included. However, the molecule-molecule interaction may be altered significantly due to interactions with the substrate, such as charge transfer or molecular deformations.\cite{hormann2019sample, hormann2022bistable} This is the case for naphthalene on Cu(111) and chemisorbed TCP on Pt(111), where the interactions change qualitatively. For this reason, it is more accurate to assume non-interacting molecules ($E_\mathrm{mol-mol} = 0$) in the first learning step.

In a second step, we learn the residual $\Delta E_\mathrm{form}$ resulting from predictions of adsorbed adlayers and the respective DFT calculations. 

\begin{equation}
	\Delta E_\mathrm{form} = E_\mathrm{form}^\mathrm{approx} - E_\mathrm{form}
\end{equation}

This two-step procedure allows (A) using large unit cells in the first step to learn molecule-molecule interactions of pseudoincommensurate adlayers and (B) learning the influence of the substrate in the second step. This enables us to consider adlayer structures with hundreds of molecules per unit cell.

\paragraph{Computational structure determination.} To determine the most energetically favorable pseudo-incommensurate structure of TCP on Pt(111) (for naphthalene on Cu(111) the structures are known form experiment\cite{yamada2010novel, forker2014complex}), we perform global optimization using a Metropolis algorithm. This approach is reminiscent of the simulated annealing optimization we used in a previous publication\cite{hormann2019sample}. For the current work, the Metropolis algorithm allows a more efficient sampling of structures containing up to \num{400} molecules per unit cell: The Metropolis algorithm chooses a new pseudo-incommensurate structure (with new lattice parameters) and directly evaluates its formation energy $E_\mathrm{form}(x_{i+1})$. The new structure is accepted based on the following equation, where $E_\mathrm{form}(x_{i})$ is the energy of the previous structure:

\begin{equation}
    p = \min \left( 1, \exp \left( \frac{E_\mathrm{form}(x_{i}) - E_\mathrm{form}(x_{i+1})}{k_\mathrm{B} T} \right) \right)
\end{equation}

Unlike our earlier simulated annealing optimization, the Metropolis algorithm does not perform full local geometry optimization on each pseudo-incommensurate structure. Forgoing these optimizations is merited in the case of TCP on Pt(111), since the energy gained through geometry optimization is in the meV-range, i.e. pseudo-incommensurate structures of TCP on Pt(111) are already energetically and geometrically very close to the respective optimized structure. Based on the results of the Metropolis algorithm, we optimize the most energetically favorable pseudo-incommensurate structures. Thereby we optimize the $x$, $y$, $z$ coordinate and orientation around the $z$ axis of each of the hundreds of molecules in the structure using a Broyden-Fletcher-Goldfarb-Shanno algorithm.\cite{broyden1970convergence, fletcher1970new, goldfarb1970family, shanno1970conditioning}

\paragraph{Density functional theory calculations.} Computational data is reused from previous work: (A) Data for naphthalene on Cu(111) comes from \textit{SAMPLE: Surface structure search enabled by coarse-graining and statistical learning}\cite{hormann2019sample} and (B) data for TCP on Pt(111) is taken from \textit{From a bistable adsorbate to a switchable interface: tetrachloropyrazine on Pt (111)}.\cite{hormann2022bistable} The data are openly available in the NOMAD repository at \href{https://dx.doi.org/10.17172/NOMAD/2023.05.01-1}{doi:10.17172/NOMAD/2023.05.01-1} and \href{https://dx.doi.org/10.17172/NOMAD/2022.03.15-1}{doi:/10.17172/NOMAD/2022.03.15-1}. Information about convergence tests and computational settings can be found in our earlier publications.

\section{Results and discussion}

As discussed in the introduction, structural superlubricity relies on incommensurability of the interface. Whether commensurate or incommensurate structures form is governed by the balance of molecule-substrate interactions and molecule-molecule interactions.\cite{hooks2001epitaxy, mannsfeld2005understanding} Strong molecule-molecule interactions and a comparatively weak corrugation of molecule-substrate interactions allow maximising the energy gain from interactions between molecules. This will most likely lead to incommensurate layers. Conversely, a large corrugation of molecule-substrate interactions and comparatively weak molecule-molecule interactions force the molecules to remain in energetically favorable adsorption sites. Any energy gain from favorable molecule-molecule interactions would be outweighed by the energy penalty from unfavorable molecule-substrate interactions. This leads to commensurate layers. Concurrently, in an incommensurate adlayer, the aforementioned delicate balance of interactions may change locally depending on the respective positions of adsorbate atoms and atoms of the molecule. This may lead to the formation of static distortion waves\cite{novaco1977orientational, novaco1979theory, mctague1979substrate, novaco1980theory, meissner2016flexible, forker2017classification} which we expect to result in a breakdown of structural superlubricity.

\paragraph{Structures of continuous adlayers.} To determine the static friction coefficient of naphthalene on Cu(111) and TCP on Pt(111) we must first know the interface structure(s) of both systems. Therefore, we identify the most energetically favorable structures of both materials. Thereby we consider both commensurate (one molecule per unit cell) and pseudo-incommensurate structures. Truly incommensurate interfaces contain an infinite number of molecules per unit cell. Since each molecule has a different adsorption site the friction coefficient becomes zero in this case. However, the electronic structure of an infinite number of molecules is impossible to calculate with first principles methods. Therefore, we approximate incommensurate with large but finite (pseudo-incommensurate) unit cells that will exhibit a finite friction coefficient.

\begin{figure}[h!]
	\centering
	\includegraphics[width=\linewidth]{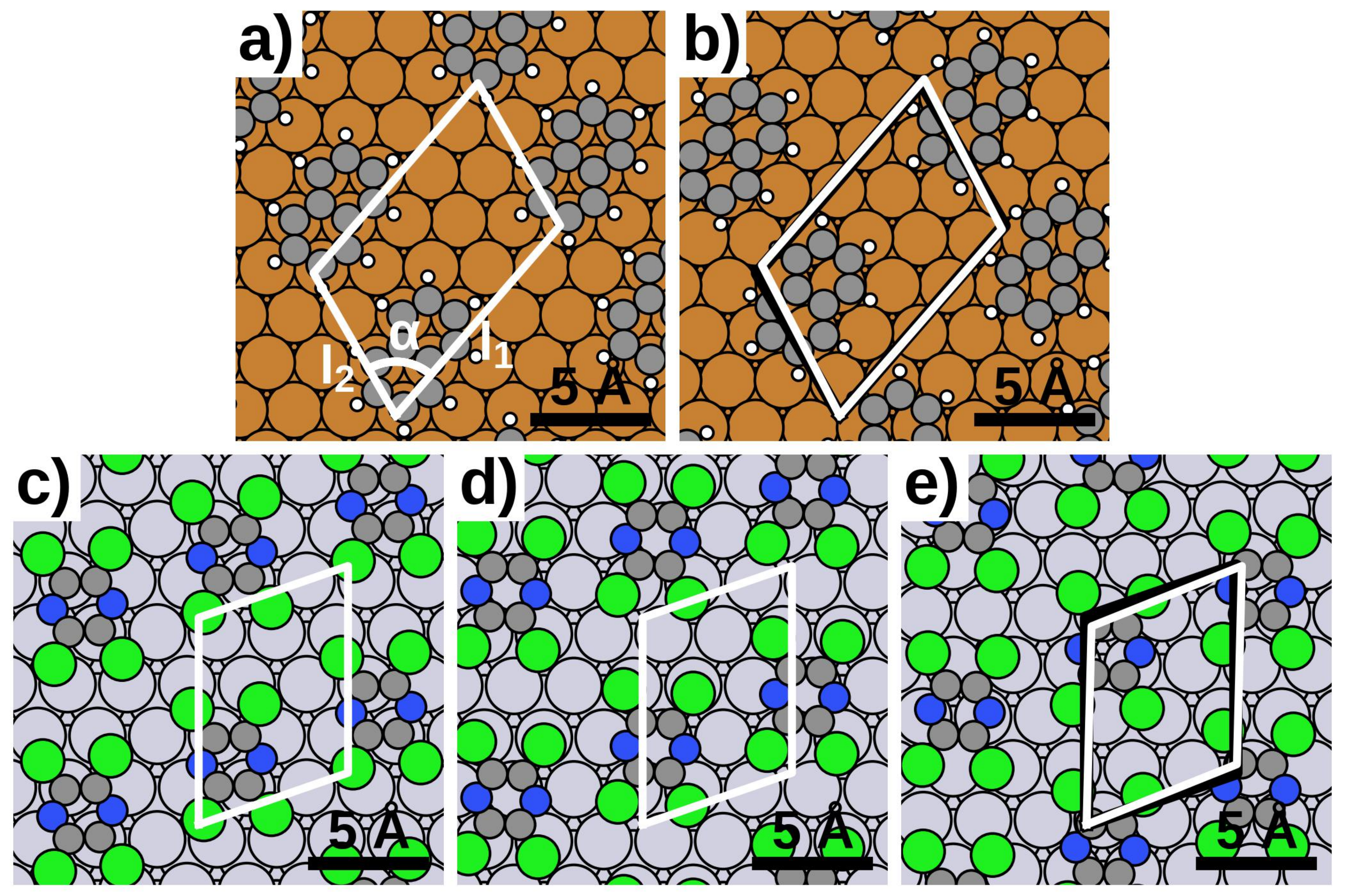}
	\caption{Close-packed adlayer structures used to calculate static friction coefficients; a) commensurate structure of naphthalene on Cu(111); b) incommensurate structure of naphthalene on Cu(111); c) commensurate structure of chemisorbed TCP on Pt(111); d) commensurate structure of physisorbed TCP on Cu(111); e) incommensurate structure of physisorbed TCP on Cu(111); unit cells are shown in white, for in incommensurate structures (b, e) the unit cells of the respective commensurate structures (a, d) are shown in black; Lattice parameters are given below:}
\begin{tabular}{c|ccc}
    & $l_1~/~\AA$ & $l_2~/~\AA$ & $\alpha~/~^{\circ}$ \\\hline
    a) & 11.7 & 7.6 & 70.9 \\
    b) & 11.5 & 7.9 & 68.4 \\
    c) & 7.3 & 9.6 & 70.9 \\
    d) & 7.3 & 9.6 & 70.9 \\
    e) & 7.5 & 9.2 & 66.7 \\
\end{tabular}
	\label{fig:geometries}
\end{figure}

In the case of naphthalene, we can make use of experimentally determined commensurate and incommensurate structures.\cite{yamada2010novel, forker2014complex} Both structures exhibit a similar molecular orientation and only have slightly different molecular periodicities. We approximate the incommensurate structure with a pseudo-incommensurate structure containing \num{150} molecules per unit cell (see figure \ref{fig:geometries}b). The commensurate structure contains \num{1} molecule per unit cell (see figure \ref{fig:geometries}a).

In the case of TCP on Pt(111), we are not aware of an experimental structure determination. Therefore, we use theoretically determined structures based on the results from an earlier publication.\cite{hormann2022bistable} We select structures based on the most favorable energy per molecule. The energy per molecule is the appropriate measure since a sub-monolayer coverage is a good assumption in nanoscale friction experiments. To determine the energies and forces of all structures (also those of naphthalene on Cu(111)), we use an MLIP. As an improvement over the MLIP used in our previous publication\cite{hormann2022bistable} we now employ a descriptor based on SOAP\cite{bartok2013representing} instead of radial distance functions. For the sake of consistency, we recalculate the formation energy for all structures. Details and a comparison between the MLIP and the old machine-leaning models are given in the Methods Section and the Supporting Information. 

As stated above, TCP forms chemisorbed and physisorbed structures on Pt(111). For the present study, we use the physisorbed structure with the most favorable formation energy per molecule, which is pseudo-incommensurate and contains \num{240} molecule per unit cell (see figure \ref{fig:geometries}e. We obtain this structure by using a Metropolis-algorithm for global minimum search. This search reconfirms the findings of our earlier study\cite{hormann2022bistable}, which has strongly indicated that the most favorable physisorbed structure is incommensurate. Additionally, we include a chemisorbed and physisorbed commensurate structure, both containing \num{1} molecule per unit cell as reference points in our study. The unit cells of these commensurate structures constitute the closest possible match compared to the one-molecule unit cell of the pseudo-incommensurate structure (see figure \ref{fig:geometries}c and d. We note that our previous study\cite{hormann2022bistable} has shown that the most favorable chemisorbed structure of TCP on Pt(111) is commensurate.

\begin{figure}[h!]
	\centering
	\includegraphics[width=\linewidth]{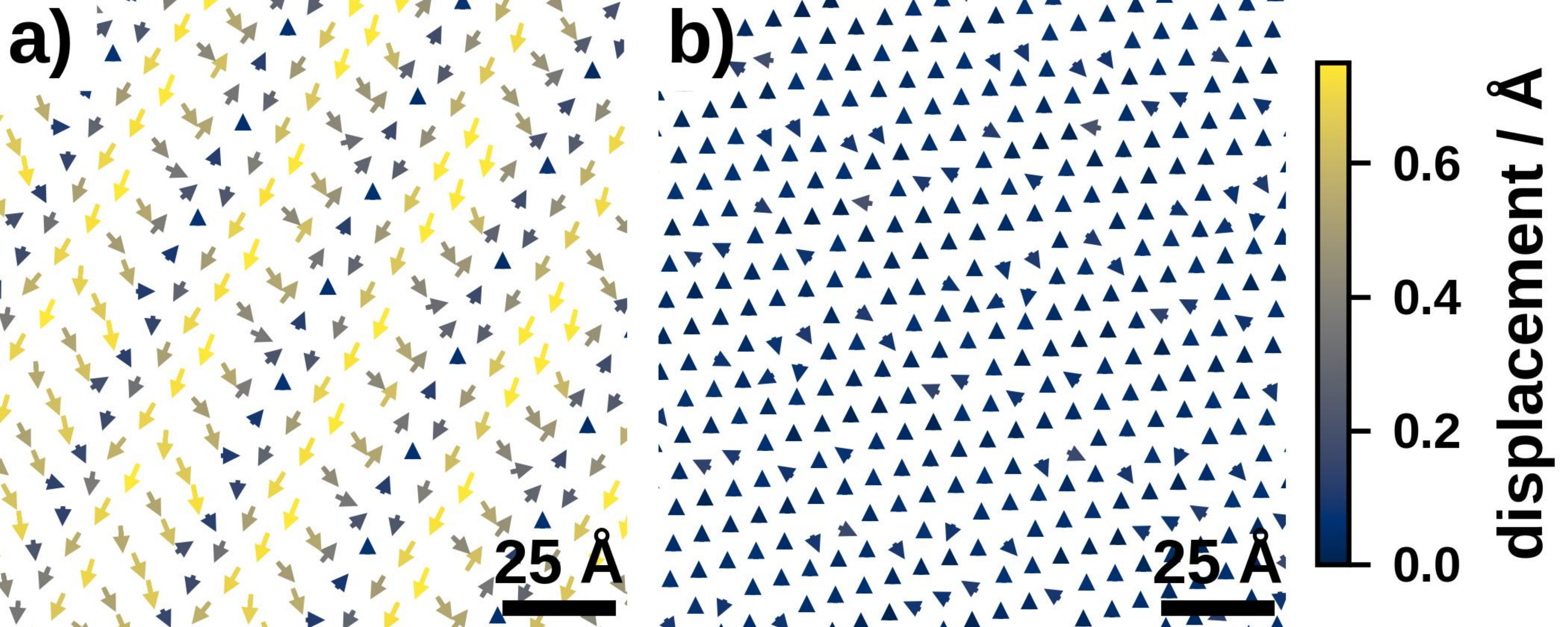}
	\caption{Displacement vectors between the molecules of the pseudo-incommensurate and optimized structures of a) naphthalene on Cu(111) and b) physisorbed TCP on Pt(111); displacement vectors are scaled by a factor of \num{5} and color-coded for better visualization.}
	\label{fig:static_distortion_wave}
\end{figure}

So far we have only considered structures where all molecules are neatly arranged on a grid. However, in the (pseudo-)incommensurate structures the molecules assume a great variety of different adsorption sites -- some of which may be located on energetically unfavorable positions of the PES (see molecule-substrate interactions). Such molecules may move to assume more energetically favorable adsorption sites. Naturally, this will incur an energy penalty from unfavorable molecule-molecule interactions. Therefore, such deformations will only occur to the extent where the energy gains outweigh the penalties. We expect that deformations of a significant magnitude, such as static distortion waves, have a significant impact on the frictional properties of the interface. To investigate the occurrence of such deformations, we use our MLIPs to conduct geometry optimizations (of the positions and orientations of the molecules) of all molecules in the pseudo-incommensurate structures (the lattice vectors remain fixed). We will hereafter refer to these optimized pseudo-incommensurate structures as optimized structures. The structure ofnaphthalene on Cu(111) deforms significantly during the geometry optimization and forms a static distortion wave, as shown by the Moir{\'e} pattern in figure \ref{fig:static_distortion_wave}. Such Moir{\'e} patterns were observed in an experimental study\cite{forker2014complex} before, albeit with a different stripe distance. This difference likely results from the sensitive dependence of Moir{\'e} patterns on the lattice parameters which may differ between theory and experiment due to the approximation via pseudo-incommensurate structures and the uncertainty of the underlying DFT calculations. Conversely, the pseudo-incommensurate structure of physisorbed TCP on Pt(111) remains largely unchanged. The reasons for these behaviors will be discussed in the next section.

\paragraph{Interactions of the molecules on the surface.} To understand why naphthalene on Cu(111) forms static distortion waves and why TCP on Pt(111) does not, it is instructive to analyze the different interactions that determine the stability of an adlayer structure. The stability of adlayer structures on the metal substrate is determined by their formation energy, which we define as the difference between the combined interface system and the interface components (see Methods Section). The formation energy can be decomposed into molecule-substrate and the molecule-molecule interactions,\cite{hormann2019sample, mannsfeld2005understanding, hooks2001epitaxy} which we will analyze in the next two subsections.

\begin{figure*}[h!]
	\centering
	\includegraphics[width=1\linewidth]{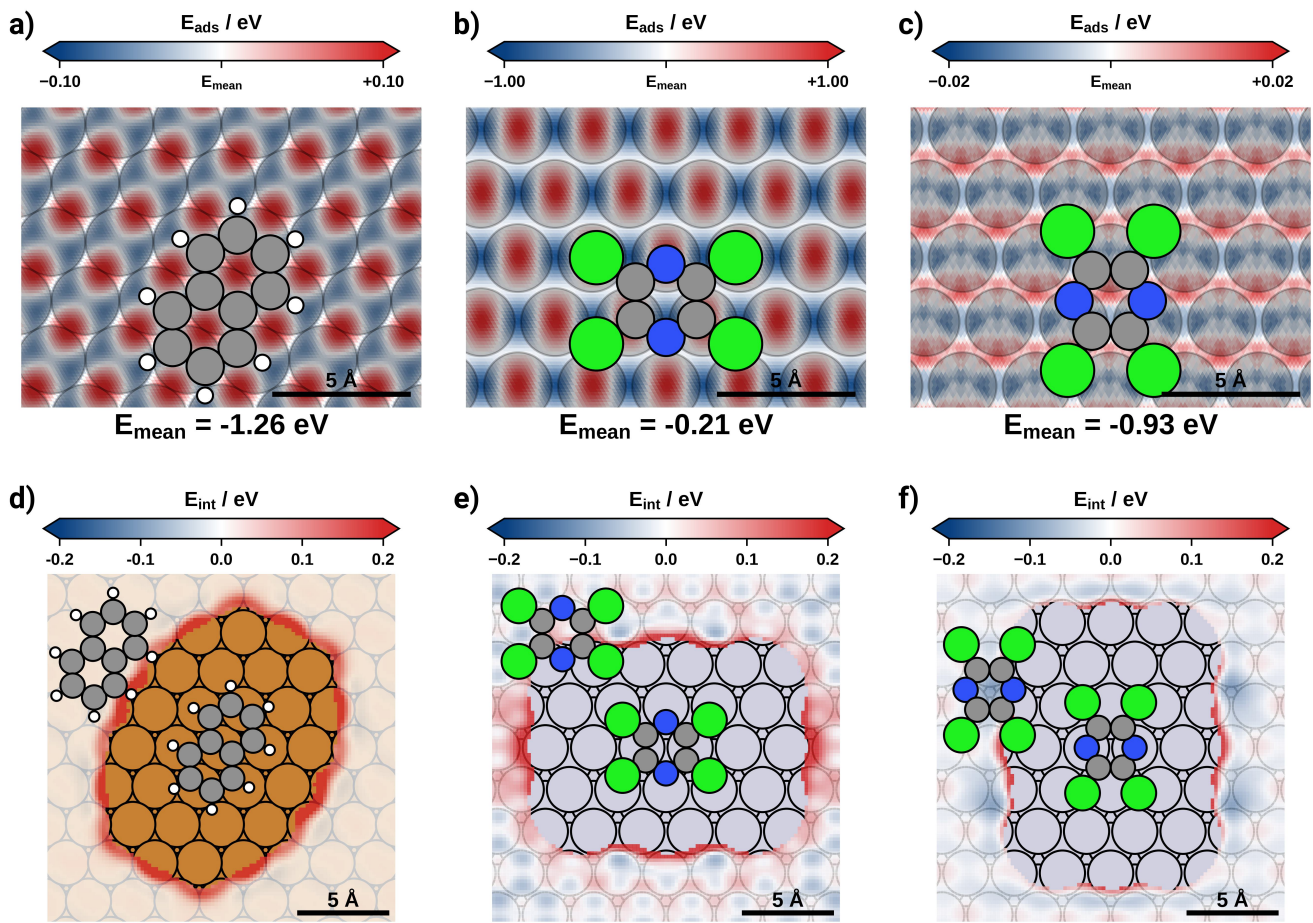}
	\caption{PES of the molecule-substrate and molecule-molecule interactions; panels a, b and c show the PES of molecule-substrate of individual molecules; the PESs shown represent 2-dimensional (x,y) cross sections of the 4-dimensional (naphthalene and physisorbed TCP) or the 5-dimensional (chemisorbed TCP) PES; the molecular orientations match the orientation of the molecules in the energetically most favorable close-packed adlayers and the energetically most favorable coordinate is chosen; energies shown are relative to the  indicated at the bottom of each subfigure; for clarity, the first layer of the substrate is shown; (a) PES of naphthalene; (b) PES of chemisorbed TCP; and (c) PES of physisorbed TCP;  panels d, e and f show the PES of the molecule-molecule interactions; (d) naphthalene on Cu(111); (e) TCP in the chemisorbed state; and (f) TCP in the physisorbed state.}
	\label{fig:figure1}
\end{figure*}

\underline{Molecule-substrate interactions.} The molecule-substrate interactions can be described by the PES of an isolated molecule on the substrate. The PES not only yields insight into the formation of possible adlayer structures, but it contains a good approximation of the potential barrier an individual molecule has to overcome during interfacial sliding. We determine the PES using the MLIP. Thereby we coarse-grain the PES (see Methods Section) and concentrate on the most important degrees of freedom: These are the $x$, $y$ and $z$ coordinate of the center of mass of the molecule, as well as orientation (around the $z$ axis) and the bending (only in the case of chemisorbed TCP) of the molecule. The algorithm uses DFT-calculated energies of \SI{50} to \SI{80} adsorption geometries as input and interpolates between them. Details about this approach and the accuracy of the prediction can be found in the Methods Section and the Supporting Information.

Figure \ref{fig:figure1} shows the $x$ and $y$ dimensions of the PESs for naphthalene on Cu(111) as well as TCP on Pt(111). To determine this 2-dimensional cross-section of the PES we chose molecular orientations that match the orientation of the molecules in the energetically most favorable close-packed adlayers (see figure \ref{fig:geometries}. For the height $z$ and the bending, we chose the energetically most favorable coordinates. Naphthalene on Cu(111) is physisorbed and the molecular backbone remains planar. The corrugation of the molecule-substrate interaction of naphthalene on Cu(111) amounts to approximately \SI{0.2}{\electronvolt}.

TCP on Pt(111) can either be chemisorbed or physisorbed. The molecule-substrate interactions of chemisorbed TCP on Pt(111) exhibit a corrugation of approximately \SI{2.0}{\electronvolt}, as we have already shown in our previous work.\cite{hormann2022bistable} This is a result of (A) the formation of (site-specific) covalent bonds with the substrate and (B) the strong distortion of the molecular geometry. Conversely, physisorbed TCP on Pt(111) remains flat and the corrugation of the PES is only \SI{0.04}{\electronvolt}. Notably, this is one order of magnitude smaller than the corrugation of naphthalene on Cu(111). \reviewerchanges{We tentatively attribute this difference in corrugation to two factors, originating from the adsorption height on the one hand and the molecular structure on the other. First, the adsorption height of naphthalene on Cu(111) is significantly lower (varying between \SI{2.72}{\angstrom} and \SI{2.80}{\angstrom}\cite{hormann2019sample}) than for physisorbed TCP  on Pt (which varies between \SI{3.27}{\angstrom} and \SI{3.31}{\angstrom}\cite{hormann2022bistable}). In a recent publication, we found the corrugation of the potential energy surface generally decreases notably with adsorption height.\cite{hormann2020reproducibility} Secondly, the distance between the two aromatic rings of naphthalene fits almost perfectly onto the Cu(111) lattice constant. This allows both rings to simultaneously adopt an “ideal” position on the surface, and results in larger energy penalties when the molecule is moved away from it. Conversely, TCP does not fit perfectly onto the Pt(111) surface and thus exhibits no such distinguished adsorption position.}

\underline{Molecule-molecule interactions.} As mentioned earlier, the formation of a given interface structure is governed by a balance between the molecule-substrate and the molecule-molecule interactions. To determine the molecule-molecule interactions, we place a pair of molecules (in isolation) onto the surface. The first molecule is kept fixed while the second one is placed at different $x$ and $y$ positions neighboring the first molecule. For the height $z$ we choose the mean adsorption height and molecular orientations reflecting the orientation of the molecules in the energetically most favorable close-packed adlayers (similar to figure \ref{fig:figure1}). For each position of the second molecule, we determine the formation energy and subtract the molecule-substrate interaction of both molecules. Figure \ref{fig:figure1} shows the molecule-molecule interactions for naphthalene on Cu(111) and TCP on Pt(111).

Naphthalene on Cu(111) exhibits exclusively repulsive molecule-molecule interactions, which become nearly zero outside the zone of Pauli-pushback. If the H-atoms of two molecules are further apart then \SI{1.0}{\angstrom}, the interaction energies are equal to or smaller than \SI{0.2}{\electronvolt}. Adlayer structures with more closely packed molecules are thus energetically unfavorable. The fact that naphthalene has small and uniform molecule-molecule interactions would commonly indicate that its adlayers are commensurate. However, whether a commensurate or an incommensurate adlayer forms depends on the relative strength of the molecule-substrate and the molecule-molecule interactions.\reviewerchanges{\cite{hooks2001epitaxy, mannsfeld2005understanding}} The molecule-substrate interactions have an equally weak corrugation of approximately \SI{0.2}{\electronvolt}. This explains why naphthalene adlayers form commensurate as well as incommensurate structures in experiment.\cite{forker2014complex, yamada2010novel} Chemisorbed TCP on Pt(111) also exhibits largely repulsive molecule-molecule interactions (Figure \ref{fig:figure1}b). The energies lie within a range of \SI{-0.2}{\electronvolt} to \SI{0.2}{\electronvolt}. Since the molecule-substrate interactions of chemisorbed TCP feature a corrugation of \SI{2.0}{\electronvolt} that is approximately an order of magnitude larger than the molecule-molecule interactions, chemisorbed adlayers are commensurate, as shown in our previous publication.\cite{hormann2022bistable} Physisorbed TCP on Pt(111) shows regions of repulsive and attractive molecule-molecule interactions. The interaction energies are comparable in size to those of chemisorbed TCP (\SI{-0.2}{\electronvolt} to \SI{+0.2}{\electronvolt}). Notably, attractive molecule-molecule interactions of approximately \SI{0.1}{\electronvolt} can be observed when Cl- and N-atoms are in close proximity (this configuration is indicated in figure \ref{fig:figure1}f). These attractive molecule-molecule interactions in conjunction with the very small corrugation of the molecule-substrate interactions likely lead to a stabilization of (pseudo-)incommensurate adlayers.

\paragraph{The connection of superlubricity and incommensurability.} In discussions about structural superlubricity, one often finds the general statement, that incommensurate structures lead to vanishing lateral forces and, hence, facilitate superlubricity.\cite{baykara2018emerging} While it is true that a perfectly incommensurate interface would lead to superlubricity, real systems (including the ones discusses here) may form static distortion waves. We gauge the impact of these distortions on the frictional properties by comparing the following structures: (A) We take commensurate structures of naphthalene on Cu(111) as well as chemisorbed and physisorbed TCP on Pt(111) as a reference point for the static friction coefficient. (B) We investigate perfectly pseudo-incommensurate (identical spacing for all molecules) structures of naphthalene on Cu(111) and physisorbed TCP on Pt(111) to determine a lower boundary for the friction coefficients. These pseudo-incommensurate structures closely match the commensurate structures to allow a direct comparison. (C) We analyze optimized structures of naphthalene on Cu(111) and physisorbed TCP on Pt(111), that is pseudo-incommensurate structures whose molecular $x$, $y$ and $z$ coordinates as well as orientations were allowed to relax and to form -- in case of naphthalene -- static distortion waves.

\begin{figure*}[h!]
	\centering
	\includegraphics[width=\linewidth]{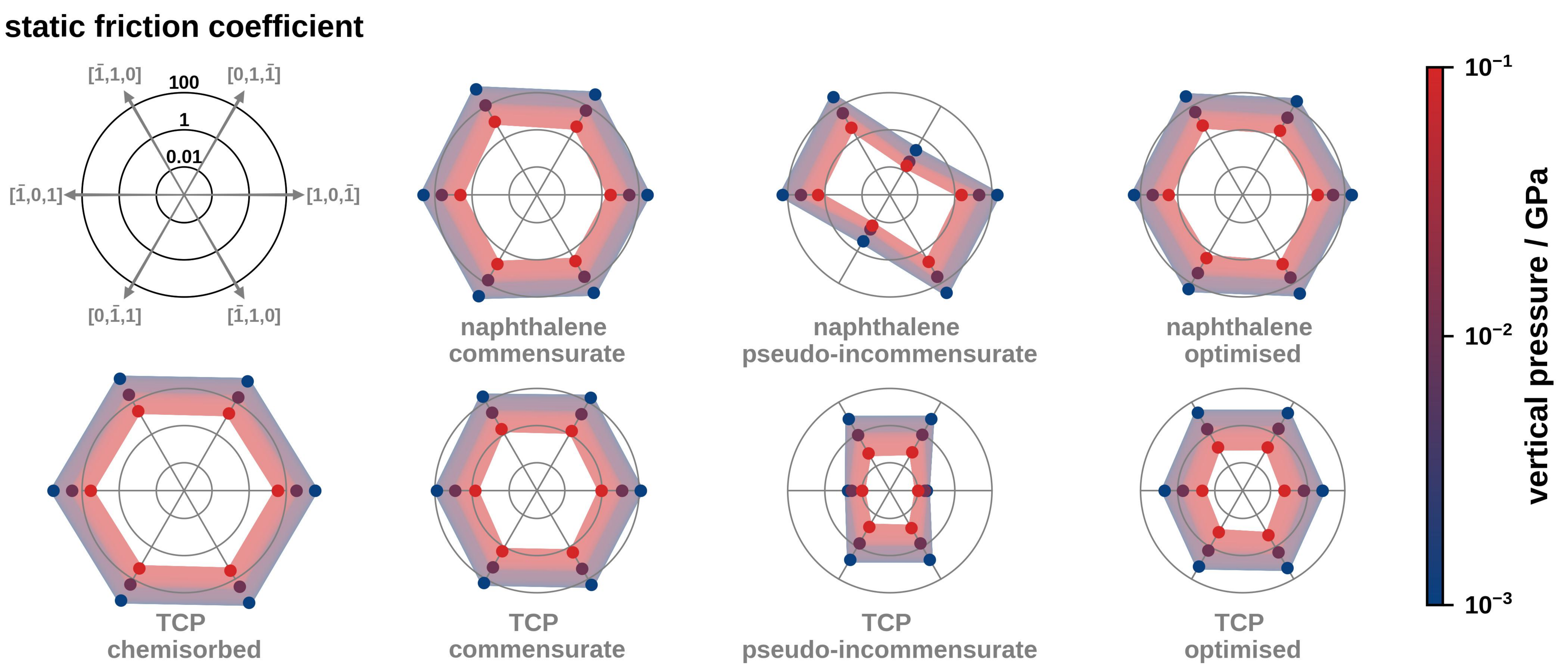}
	\caption{Dependence of the static friction coefficients for different structures of naphthalene on Cu(111) and TCP on Pt(111) on the direction of displacement; The solid dots in the plot indicate calculated static friction coefficients while the shaded area is an interpolation which should serve as a guide for the eye.}
	\label{fig:friction_coefficient_direction}
\end{figure*}

To determine the static friction coefficient, we shift the different molecular adlayers across the surface. The paths are given by the primitive substrate lattice vector of the (111)-surface. The different directions (hereafter called primitive directions) are indicated in figure \ref{fig:friction_coefficient_direction}. We note in passing that opposite directions appear symmetric due to substrate symmetries, but are strictly not symmetric due to the adsorbate layer. At equidistant points along these paths, we determine adsorption energies and forces acting on the adlayer using the MLIP. Thereby we apply a normal force individually to all molecules in the unit cell. To find the equilibrium adsorption height of each molecule the $z$ coordinate of its center of mass is optimized (under the impact of the vertical force) using the MLIP. \reviewerchanges{We note in passing that the MLIP determines forces by taking the (numerical) derivative of the interfacial PES $E_\mathrm{form}$. For instance, the lateral force $F_\mathrm{lat}$ is determined using $F_\mathrm{lat} = - \nabla E_\mathrm{form} \cdot \vec{s}$, where $\vec{s}$ is the sliding direction.This approach is commonly used in literature.\cite{zhong1990first, zhang2023interlayer}} The lateral force is a result of the interfacial PES (molecule-substrate plus molecule-molecule interactions). Its maximum value opposite the shear direction (i.e. when moving ``up'' an energy barrier) directly relates to the static friction coefficient. The friction coefficient can be determined from the relationship of the lateral force \reviewerchanges{$F_\mathrm{lat}$} and the respective normal force $F_\mathrm{z}$:

\reviewerchanges{
\begin{equation}
    \label{eq:friction_coefficient}
	\mu_\mathrm{s} = \frac{1}{F_\mathrm{z}} \max(F_\mathrm{lat})
\end{equation}
}

\begin{figure*}[h!]
	\centering
	\includegraphics[width=\linewidth]{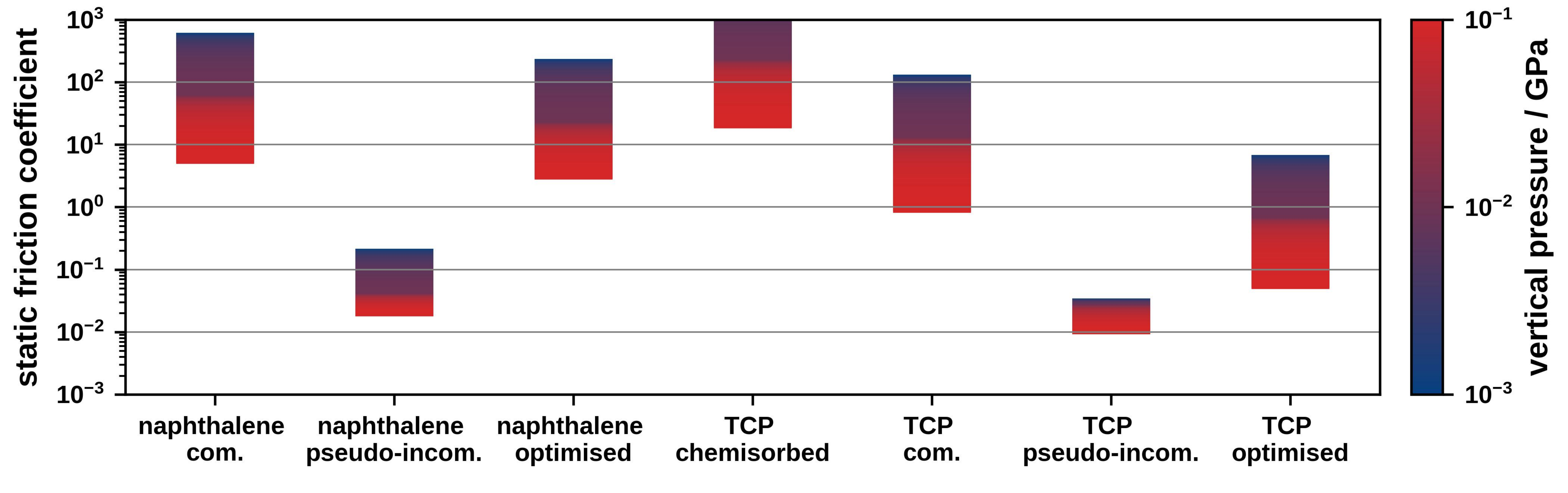}
	\caption{Comparison of static friction coefficients for different structures of naphthalene on Cu(111) and TCP on Pt(111).}
	\label{fig:friction_coefficient_comparison}
\end{figure*}

Figures \ref{fig:friction_coefficient_direction} and \ref{fig:friction_coefficient_comparison} present a comparison of the static friction coefficients of naphthalene and TCP respectively. Thereby we compare commensurate, pseudo-incommensurate and optimized structures. \reviewerchanges{For a clear representation in figures \ref{fig:friction_coefficient_direction} and \ref{fig:friction_coefficient_comparison} we convert the vertical forces into vertical pressures. The range of vertical pressures from \SI{0.001}{\giga\pascal} to \SI{0.1}{\giga\pascal} reflects experimentally used pressures.\cite{baykara2018emerging, liu2017robust}}

Commensurate naphthalene on Cu(111) exhibits large static friction coefficients of approximately \num[retain-zero-exponent]{5e0} to \num{5e2} that are isotropic in all $6$ primitive directions of the 111-surface (see figure \ref{fig:friction_coefficient_comparison}). Conversely, the pseudo-incommensurate adlayer of naphthalene displays a strongly anisotropic static friction coefficient. In $[0,1,\bar{1}]$- and $[0,\bar{1},1]$-direction it has a static friction coefficient of approximately \num{2e-2} to \num{5e-1} that is nearly $3$ orders of magnitude smaller than that of the commensurate adlayer. In the other directions, the friction coefficients are comparable to the commensurate naphthalene structure. \reviewerchanges{The reason for this is twofold: First, the pseudo-incommensurate structure of naphthalene has its largest extent in the $[0,1,\bar{1}]$/$[0,\bar{1},1]$direction (see figure S15 in the Supporting Information). Therefore, the lattice mismatch in this direction is maximal, which results in a minimal static friction coefficient. Second, the PES of the molecule-substrate interaction (see figure \ref{fig:figure1}a) exhibits ``grooves'' in the $[0,1,\bar{1}]$/$[0,\bar{1},1]$-direction. The molecule can move along these ``grooves'' without encountering large barriers. In the other primitive directions, the barriers are higher, leading to a larger static friction coefficient.} \reviewerchanges{Finally, the optimized structure exhibits a static friction coefficient comparable to the commensurate adlayer. This behavior is a result of the formation of static distortion waves, where the molecules adjust their positions (and orientations) to benefit from favorable molecule-substrate interactions (see figure \ref{fig:figure1}). Hence, molecules are generally close to the local minima of PES, similar to a commensurate structure. When the molecular adlayer is slid over the substrate, the lateral forces acting on individual molecules point in the same direction, which leads to a larger friction coefficient.}

Chemisorbed TCP on Pt(111) has a very large static friction coefficient, with lateral forces being larger than the vertical force. We note that we did not allow for the sliding of the atoms of the metal substrate, which may in the case of chemisorbed TCP offer less resistance to lateral movement. Commensurate physisorbed adlayers show significantly lower static friction coefficients of approximately \num[retain-zero-exponent]{0.9e0} to \num{1.2e2}, which are $2$ orders of magnitude smaller than in the chemisorbed adlayer. In both cases, the friction coefficients are largely isotropic. Notably, commensurate adlayers of TCP have a static friction coefficient that is half an order of magnitude smaller than that of commensurate adlayers of naphthalene. This is a result of the small corrugation of the molecule-substrate interaction, which is $1$ order of magnitude smaller than that of naphthalene on Cu(111). The pseudo-incommensurate adlayer of physisorbed TCP exhibits anisotropic static friction coefficients, with the minimum of \num{1.0e-2} to \num{0.3e-1} being found in the $[1,0,\bar{1}]$- and $[\bar{1},0,1]$-directions. \reviewerchanges{We attribute this anisotropy to, (A) ``grooves'' in the PES (see figure \ref{fig:figure1}c) that allow the molecule to move in the $[1,0,\bar{1}]$- and $[\bar{1},0,1]$-directions without encountering large barriers and (B) a maximal lattice mismatch in the respective primitive directions.} These results are comparable to those of a pseudo-incommensurate adlayer of naphthalene. Compared to that the static friction coefficient of the optimized structure is minimally larger (\num{5.0e-2} to \num[retain-zero-exponent]{6.0e0}) and the directional dependence becomes isotropic. This friction coefficient is $2$ orders of magnitude smaller than that of the commensurate adlayer of physisorbed TCP. \reviewerchanges{Importantly, when comparing physisorbed TCP and naphthalene we find that the static friction coefficients exhibit a significantly different behavior when optimizing the respective pseudo-incommensurate structures: In the case of physisorbed TCP the friction coefficient increases by half an order of magnitude, whereas, in the case of naphthalene the friction coefficient increases two orders of magnitude.} This is because the formation of static distortion waves in pseudo-incommensurate adlayers of TCP is suppressed by attractive molecule-molecule interactions, in contrast to naphthalene adlayers, where molecule-substrate interactions are dominant.


We note in passing that the static friction coefficients for all structures exhibit a considerable dependence on the vertical pressure. This has two reasons: First, friction coefficients are calculated using equation \ref{eq:friction_coefficient}, where the vertical force (which results from the vertical pressure) is in the denominator. Hence, in the case of similar lateral forces, a larger vertical force leads to a smaller friction coefficient. Second, a larger vertical pressure ``pushes'' the molecular adlayer closer to the substrate. In an earlier work\cite{hormann2020reproducibility} we have shown that the corrugation of the interfacial PES increases when the adsorption height decreases.

In summary, we find a strong impact of the type of commensurability on the static friction coefficient. As expected, pseudo-incommensurate structures exhibit ultra-low friction. Lower friction coefficients could in principle be computationally obtained with structures containing even more molecules per unit cell, but we do not expect that larger unit cell sizes impact the likelihood of the formation of static distortion waves. Static distortion waves cause a significant increase in friction. The formation of incommensurate organic/metal interfaces requires a physisorbed state since chemisorbed molecules are usually strongly bonded to a specific site on the substrate. Moreover, static distortion waves may occur if the molecule-molecule interactions are small compared to the molecule-substrate interaction, as is the case for naphthalene on Cu(111). The stabilization of an incommensurate adlayer requires strongly corrugated molecule-molecule interactions that leads to distinguished molecular arrangement, as is the case in physisorbed TCP.

\section{Conclusion}

In conclusion, we quantify the dependence of the static friction coefficient on the type of commensurability and in particular on the formation of static distortion waves. This requires the determination of the energetically most favorable interface structures, which depend on a balance of molecule-molecule interactions and molecule-substrate interactions. Because these interactions are governed by quantum mechanical effects such as interfacial charge transfer, we conduct our simulations using DFT-level accuracy. This precision allows us to model the formation static distortion waves. However, static distortion waves are a mesoscale structural phenomenon, which necessitates system sizes that are intractable with DFT alone. We overcome this challenge by developing an MLIP capable of calculating energies and forces of interface structures containing hundreds of molecules per unit cell. Using this MLIP, we investigate the frictional properties of two exemplary organic/metal interfaces: naphthalene on Cu(111) and TCP on Pt(111). Naphthalene on Cu(111) is physisorbed while TCP on Pt(111) can be either physisorbed or chemisorbed. We compare static friction coefficients of (A) commensurate, (B) pseudo-incommensurate and (C) geometry-optimized pseudo-incommensurate adlayer structures of both interfaces. We find that the optimized pseudo-incommensurate structure of naphthalene on Cu(111) exhibits static distortion waves leading to a static friction coefficient of similar magnitude as in commensurate structures. This is a result of dominant molecule-substrate interactions compared to the molecule-molecule interactions. Conversely, pseudo-incommensurate structures of TCP on Pt(111) are stabilized by strongly corrugated molecule-molecule interactions. Therefore, the pseudo-incommensurate structure of TCP on Pt(111) displays a static friction coefficient that is approximately $2$ orders of magnitude smaller than that of naphthalene on Cu(111). Based on these results we can formulate design principles to produce organic/metal interface systems that display superlubricity: (A) The formation of incommensurate structures requires relatively uniform molecule-substrate interactions, which are commonly only found in physisorbed systems. (B) Perfectly incommensurate structures, which (potentially) exhibit superlubricity, are geometrically stabilized in systems with strongly corrugated molecule-molecule interactions. On this basis, we can state that in organic/inorganic interface systems, incommensurability implies ultra-low friction only in systems where dominant molecule-molecule interactions suppress the formation of static distortion waves.

\section{Supporting Information}

The Supporting Information contains details on the MLIPs used for the calculations of static friction coefficients in this work. We explain the construction of the MLIPs and test their accuracy. Additionally, we provide further analysis of the molecule-molecule interactions and convergence test for the unit cell size of the pseudo-incommensurate structures.

\section{Data availability}

The data supporting this research are openly available in the NOMAD repository at \href{https://dx.doi.org/10.17172/NOMAD/2023.05.01-1}{doi:10.17172/NOMAD/2023.05.01-1} and \href{https://dx.doi.org/10.17172/NOMAD/2022.03.15-1}{doi:/10.17172/NOMAD/2022.03.15-1}.

\section{Acknowledgements}

Financial support by the UFO grant of the Land Steiermark via the project ``Schaltbare Superschmierfähigkeit'' is gratefully acknowledged by LH. JJC and OTH gratefully acknowledge financial support from the Austrian Science Fund (FWF) via the project Y1157-N36 ``MAP-DESIGN''. Computational results have been achieved using the Vienna Scientific Cluster (VSC).

\section{Author contributions statement}


\textbf{Lukas Hörmann:} Conceptualization, Methodology, Software, Validation, Formal analysis, Investigation, Data curation, Funding acquisition, Writing – original draft, Writing – review \& editing. 
\textbf{Johannes J. Cartus:} Investigation, Formal analysis, Writing – review \& editing.
\textbf{Oliver T. Hofmann:} Conceptualization, Resources, Investigation, Project administration, Funding acquisition, Supervision, Writing – review \& editing.

\section*{Conflicts of interest}

The authors declare no competing interests.

\bibliography{bibliography}

\end{document}